\documentstyle[epsf]{l-aa}     
\epsfverbosetrue
  \newcommand{\Msolar}{\mbox{\,$\rm M_{\odot}$}}        
  \newcommand{\Rsolar}{\mbox{\,$\rm R_{\odot}$}}        
  \newcommand{\Lsolar}{\mbox{\,$\rm L_{\odot}$}}        
\newcount\equanumber      \equanumber=0
{\global\advance\equanumber by 1}
\def\begineq#1{\global\advance\equanumber by 1
              \begingroup $$ #1 \hfill \eqno {\rm (\number\equanumber)}}

\newcommand{\gppr}{\stackrel{>}{\scriptstyle \sim}}
\newcommand{\gappr}{\raisebox{-0.4ex}{$\gppr $}}
\def\Ang{{\rm\AA}}

\def\ion#1#2{#1\,{\sc #2}}
\def\1640{\ion{He}{ii}$\;\lambda1640$}

\def\Teff{\hbox{$T_{\rm eff}$}}
\def\tief #1{_{\rm #1}}

\def\Nh{\hbox{$N_{\rm H}$}}
\newcommand{\Mdot}{\stackrel{.}{M}}

\def\AaA #1{A\&A #1}
\def\Supp #1{A\&AS #1}

\begin{document}
  
\thesaurus{ 08 (08.02.5; 08.06.3; 08.09.2: SMC3; 08.09.2: RX~J0048.4--7332; 
	08.13.2; 13.25.5) }

\title{Extragalactic symbiotic systems\thanks{Based on data collected 
				with the ROSAT and the IUE satellites}}
\subtitle{IV. The supersoft X-ray source SMC~3}
  \author{Stefan Jordan\inst{1}, Werner Schmutz\inst{2}, Burkhard Wolff\,\inst{1}, 
		Klaus Werner\inst{1}\inst{3}, Urs M\"urset\inst{2}}

   \offprints{S. Jordan}

  \institute{    
	Institut f\"ur Astronomie und Astrophysik,
               Universit\"at Kiel, 
                 D-24098 Kiel, Germany
\newline                    
        E-mail: jordan/wolff/werner@astrophysik.uni-kiel.d400.de
	\and
	Institut f\"ur Astronomie, ETH Zentrum, CH-8092
                Z\"urich, Switzerland;
        E-mail: schmutz/muerset@astro.phys.ethz.ch
\and 
        Lehrstuhl Astrophysik, Universit\"at Potsdam, 
        Am Neuen Palais 10, D-14469 Potsdam, Germany
}

    \date{Received, accepted}

    \maketitle
\markboth{Jordan et al.: The supersoft symbiotic nova SMC3}{xx}

\begin{abstract}
We present a consistent model for the UV and supersoft X-ray emission
from the symbiotic nova SMC3 (=~RX~J0048.4--7332). Following the present 
picture of symbiotic stars, the 
model consists of radiation from a hot star and an emission nebula excited 
by that star. The observations were compared to  theoretical models
in which the hot star's emission is calculated with the help of 
hydrostatic and Wolf-Rayet-type non-LTE  model atmospheres.
Our analysis clearly shows evidence for mass loss rates of 
several $10^{-6}$~\Msolar/yr
.
The minimum effective temperature compatible with both the observed UV and
X-ray flux is about $260\,000$~K, which is higher than in any other star
analyzed with sophisticated NLTE model atmospheres.
 Since the hydrostatic surface is hidden
by the stellar wind no upper limit for the temperature can be determined.
However, we were able to determine the total luminosity of a symbiotic
nova with reasonable accuracy ($L_{\mbox{SMC3}}=10^{4.05\pm 0.05}$\,L$_\odot$).
This value is well below the Eddington limit ($\approx50\,000$~L$_\odot$).
In order to reproduce the observed energy distribution 
a carbon-to-helium ratio $>2\cdot 10^{-4}$ --- leading to an absorption
edge at 0.39\,keV ---  is necessary. 

\keywords{ 
        Binaries: symbiotic -- 
	Stars: fundamental parameters --
	Stars: Individual: SMC3 --
	Stars: Individual: RX~J0048.4--7332 --
        Stars: mass-loss -- 
	X-rays: stars}
\end{abstract}

\section{Introduction}

ROSAT observations established  the class of `supersoft
X-ray sources' (`SSS'; e.g.\ Hasinger 1994) with almost no
flux at $h\nu\ge 0.5$~keV; normally, close-by 
low-luminosity objects like single white dwarfs or  cool coronal stars
are excluded from this definition.
The nature of many SSS is still a matter of controversial
discussion. 
RX~J0048.4--7332 is one of 
ten  SSS confirmed as members of the Magellanic Clouds
(Kahabka \& Pietsch 1993; Pietsch \& Kahabka 1993).
In this paper we investigate the hypothesis  that the 
X-ray flux of this object is due to photospheric emission of a
very hot white dwarf. Given a sufficiently high temperature
the photospheric  emission of white dwarfs  is measurable in the ROSAT window
(e.g. Jordan et al. 1994b, Wolff et al. 1996).
A hot white dwarf is hardly the correct explanation for all supersoft 
X-ray sources. However, at least for some sources, the presence of a very hot
star is also in agreement with other evidence:
For instance, in symbiotic systems there are high excitation
nebular emission lines combined with a hot continuum in the UV.
The supersoft appearance is also expected from the theoretical side;
in particular, for symbiotic novae Sion \& Starrfield (1994) predict 
phases in which a very hot white dwarf can be detected.

Indeed, the supersoft source RX~J0048.4--7332 coincides with the 
symbiotic nova SMC3
that has been in outburst since 1981 (Morgan 1992). Thus, for
this system it appears natural
to propose that a hot white dwarf is  the source of the X-ray flux.
Symbiotic stars are interacting binaries consisting of a red giant 
and a very hot white dwarf (typically $T\tief{eff}\sim100\,000$~K),
whose  radiation ionizes the emission nebula.
 In the optical spectral range we 
observe a composite spectrum from the cool star and the nebula.
The UV spectrum is dominated by the nebular emission with only
a small contribution from the hot star that becomes dominant
at the shortest wavelengths.
The spectral range best suited for a direct
observation of the hot component of a symbiotic system is the
soft X-ray and EUV part of the electromagnetic spectrum, where these
stars emit the bulk of their energy 
(e.g. RR~Tel, see Jordan et al.\ 1994a, hereafter JMW).

Vogel \& Morgan (1994; Paper I) obtained an IUE spectrum of SMC3 which 
looks typical for a symbiotic star of moderate excitation. 
On the other hand,
the optical spectrum (Morgan 1992; M\"urset et al.\ 1995 =
Paper III) reveals nebular ionization that is unusually high
for symbiotics, up to Fe$^{+9}$. Even more peculiar, however, are the X-ray
observations presented by Kahabka et al.\ (1994): Despite of
the large distance, the ROSAT count rate of SMC3 is about the
same as of  RR~Tel,  hitherto the  X-ray brightest of all
known galactic symbiotic
novae (Jordan et al.\ 1994a; hereafter JMW). Thus, the ratio of the
optical to the X-ray flux of SMC3 is  a challenge for  models of symbiotic
stars. On the other hand, the known distance of this extragalactic
system provides
a unique opportunity to investigate the energetics of interactions
and outbursts in symbiotic systems. 

Kahabka et al.\ (1994) have presented black body fits to the ROSAT data, which
resulted in an extremely unrealistic
luminosity of up to  $10^{13}$~L$_\odot$.
However, we will show that a reasonable value is obtained
if the data are compared to appropriate model atmospheres, and if we
consistently take into account the constraints from both the  X-ray and UV
regions of the electromagnetic spectrum.

In Sect.~\ref{data} we list the data used in our analysis, in Sections
\ref{atmo} and \ref{anal}
we describe the methods, the results are
discussed in Sect.~\ref{resu}, and Sect.~\ref{conclu} presents
some concluding remarks.

\section{Data\label{data}}

\subsection{ROSAT}

Pointed ROSAT PSPC observations were retrieved from the ROSAT archive
and reduced using the EXSAS software (Zimmermann et al.\ 1994). 
Kahabka et al.\ (1994) describe the data, obtained during a 20~ks exposure in
April 1992, in detail.
We binned the counts into the various  energy channels  until a signal-to-noise
ratio of five was achieved. After correcting  for dead time and vignetting
the resulting total count rate amounts to
$0.204 \pm 0.004$\,counts/sec.
The observed PSPC pulse height distribution (PHD)  is displayed in 
Figs.\ \ref{fdo} -- \ref{fwr7071}.

A follow-up observation in December 1992 is not included in the
archive by the time of writing, and could therefore not be used for this work.

\subsection{IUE}

The UV spectrum of SMC3 is described in Paper I. It was observed
with the short 
wavelength ($1200~\Ang\la\lambda\la1900~\Ang$) camera of the
International Ultraviolet Explorer satellite (IUE)
at low resolution using the large ($20^{\prime \prime}\times10^{\prime \prime}$) aperture
in April 1993. 
Given the stability of SMC3 in its present outburst stage (Morgan 1992)
we assume that there was no variability so that the ROSAT 
and the UV observation can be combined.

\subsection{Optical spectra\label{optiflux}}

Optical low resolution spectra of SMC3 have been published by
Morgan (1992) and in Paper III. The spectrum is typical 
for a high excitation s-type symbiotic. It displays
\ion{H}{i} Balmer, \ion{He}{i}, and \ion{He}{ii} emission lines,
as well as the Raman scattered \ion{O}{vi}$\;\lambda\lambda$\,6825,\,7082 
features (cf.\ Schmid 1989). 
In addition to the typical features,
[\ion{Fe}{x}]$\;\lambda$6374 is also present
which is very rarely observed in 
symbiotic spectra. Davidsen et al. (1977) report
[\ion{Fe}{x}] emission from GX1+4,
which is a symbiotic system that contains a neutron star, 
and Webster \& Allen (1975) observed [\ion{Fe}{x}] for He2--171.
The spectrum of AS 295B contains
[\ion{Fe}{x}], [\ion{Fe}{xi}], and [\ion{Fe}{xiv}] (Herbig \& Hoffleit 1975).

From the H$\beta$ luminosity and the line ratios
given by Morgan (1992)
we derive the optical line fluxes of H$\beta$ and \ion{He}{ii}$\;\lambda$4686
listed in Table~\ref{lintab}.

\begin{table}
\caption{Continuum and integrated emission line fluxes from SMC\,3 used in the
analysis. The optical line fluxes are from Morgan (1992).}
\label{lintab}
\begin{flushleft}
\begin{tabular}{ll}
\hline\noalign{\smallskip}
\ $f$(continuum $\lambda=1300$~\AA)	
	&$2.9\cdot10^{-15}$~erg~cm$^{-2}$~s$^{-1}$~\AA$^{-1}$	\\
\ $f$(\ion{He}{ii}$\;\lambda$1640)
	&$2.3\cdot10^{-13}$~erg~cm$^{-2}$~s$^{-1}$		\\
\ $f$(\ion{He}{ii}$\;\lambda$4686)
	&$3.3\cdot10^{-14}$~erg~cm$^{-2}$~s$^{-1}$		\\
\ $f$(H$\beta$ $\;\lambda$4861)
	&$4.5\cdot10^{-14}$~erg~cm$^{-2}$~s$^{-1}$		\\
\ $f$([\ion{Fe}{x}]$\;\lambda$6374)
	&$2.0\cdot10^{-14}$~erg~cm$^{-2}$~s$^{-1}$			\\
\noalign{\smallskip}\hline	
\end{tabular}
\end{flushleft}
\end{table}

\subsection{Distance and interstellar extinction\label{distext}}

Our analysis requires reddening corrected continuum and
emission line fluxes.
The extinction can be derived by comparing the flux ratio of the
\ion{He}{ii} recombination lines $\lambda$4686 and $\lambda$1640
to the theoretical value for case-B recombination
(Hummer \& Storey 1987). Although this ratio implies negligible 
extinction,
we corrected for the foreground extinction of the SMC 
($E\tief{B-V}=0\fm05$; Bessell 1991) by  using a galactic extinction curve,
 and subsequently
applying the SMC extinction curve  by  Pr\'evot et al.\  (1984) for
$E\tief{B-V}=0\fm03$, which is the average internal extinction
of the SMC (Bessell 1991).
For the distance to the  SMC we adopt $d=61$~kpc determined
by Barnes et al.\ (1993).

\section{Atmospheres\label{atmo}}

The spectroscopic analysis is based on
state-of-the-art non-LTE 
atmospheres
as required by the high temperatures involved.
They were calculated with two different codes,
one assuming plane-parallel geometry and hydrostatic
layers, the other is appropriate for extended atmospheres
with a steady state stellar wind.
Table\,\ref{wrmod}
gives an overview on the hot model atmospheres 
that have been calculated.

\subsection{Hydrostatic models}

As in the case of RR~Tel (JMW) 
we first computed
the theoretical model spectra for the hot stellar component with  line
blanketed non-LTE model atmospheres assuming  plane-parallel geometry
and hydrostatic and radiative equilibrium. 
Besides hydrogen and helium the models contain C, N, and O.
The number of non-LTE levels assigned to each species
are 5, 12, 38, 48, 48, respectively, and a total of 333 lines were explicitly
included. The characteristics of these models were described in detail by Werner
\& Heber (1991).

\subsection{Models with a stellar wind}

Plane-parallel models are justified if the thickness of the stellar atmosphere
 is small compared to
the stellar radius. This is normally
the case for hot white dwarfs. 
However, these assumptions may not be valid
in the case of a symbiotic nova, since for several of the
symbiotic novae, there is evidence for mass loss (e.g.
Nussbaumer et al. 1995, Nussbaumer \& Vogel 1995). 
Therefore, we also calculated
models with mass loss as an additional free parameter using
 the co-moving frame code for hot stars with a strong
stellar wind developed in Kiel
(Hamann \& Schmutz 1987; Wessolowski et al.\ 1988).
The non-LTE calculations include 6 levels for the hydrogen model atom;
11 levels for helium with
5 levels for \ion{He}{i} and 5 levels for \ion{He}{ii};
12 levels for carbon with 5 levels for \ion{C}{iv} and 6
levels for \ion{C}{v}; 6 levels for nitrogen (5 levels for \ion{N}{v})
and 9 levels for oxygen (5 levels for \ion{O}{vi}). 
In total 44 levels and 54 line transitions are treated
explicitly.

The free parameter of a model with an extended atmosphere due to mass loss
are $\Teff$, $R_*$, $\Mdot/v_\infty$, and $\beta$. The radius, $R_*$,
is the inner boundary of the model. This inner boundary is either where
the expansion velocity of the envelope is subsonic or, if the 
mass loss is so strong that this region is at very large optical depths,
then the inner boundary is set where the Rosseland
optical depth $\tau_{{\rm R}} \approx 100$. 
The temperature $\Teff$ yields in combination with $R_*$ the luminosity.
$v_\infty$ and $\beta$ 
specify the velocity law of the wind. We adopt a velocity law
$v(r)=v_\infty(1-R_*/r)^{\beta}$, assuming  $\beta=1$ and
$v_\infty=1000$\,km/sec (cf. section 5.2). This
type of law resembles that found empirically for O stars 
(Groenewegen \& Lamers 1989) and agrees with the law calculated
hydrodynamically for radiation driven winds of O stars
(Pauldrach et al.\ 1986; Schaerer \& Schmutz 1995).

\section{Fit procedure}
\label{sfit}\label{anal}
There are three observational constraints that have to be reproduced
by a successful model: the X-ray flux, the UV continuum in the
IUE wavelength range, and the nebular emission line \1640;
in addition, there is a high ionization line,
[\ion{Fe}{x}]$\;\lambda$6374. Since a forbidden 
transition is very sensitive to 
the unknown density of the nebula, the observation of the [\ion{Fe}{x}]
line implies only a lower limit for the hardness of the exciting
radiation, but the observed line strength does not help to 
constrain a stellar parameter.

In order to reproduce the three observations we
have adapted the following procedure. First,
we have computed a large grid of model atmospheres with different
parameters in \Teff, $\log g$, chemical composition, 
and mass loss rates (or $\Mdot/v_\infty$, respectively).

The emergent flux does not depend on the radius
of the models. This is obviously true for hydrostatic models where
the radius is not among the model parameters. But, it is also correct
for spherically extended models with mass loss, provided that the
product $\Mdot/v_\infty/R_*^{3/2}$ is kept constant (Schmutz et al.
1989). Thus, if we assume that the continuum at $\lambda=1300$~\AA\ is
dominated by the flux of the hot star, we can compute
the stellar radius for each model, knowing the flux predicted by the
model atmosphere and the observed luminosity at 1300 \AA;
at this wavelength the hot stellar emission dominates the continuum 
in all symbiotic objects whose UV spectra have been investigated by 
M\"urset et al.\ (1991).

With the absolute dimensions of the star determined
by the observed 1300 \AA\ flux for each model of the grid,
we fed the emergent
radiation to the photo-ionization code NEBEL (Nussbaumer \& Schild 1981;
Deuel 1986; Vogel 1990). For the calculation of the emission lines we 
assume that the nebula is radiation bounded in the \ion{He}{ii} region.
We have no possibility to verify this assumption. However, 
M\"urset et al. (1991) found the \ion{He}{ii} region
to be radiation bounded for almost all symbiotic systems and
only a few  exceptions being  nearly radiation bounded.
The presence of a radiation bounded nebula
is in no contradiction to the observed X-rays since 
our calculations show that an ionization bounded nebula is not 
necessarily
optically thick below 100 \AA; material from the neutral part of the
nebula   does, of course, 
contribute to the column density towards the X-ray source.
Indeed, we find hydrogen column densities
(up to $1.1 \cdot 10^{21}$\,cm$^{-2}$)
 that are much larger than
the value  we would expect from  the reddening of SMC3
(the formula by Groenewegen \&\ Lamers 1989 yields about
$N\tief{H}=1.9 \cdot 10^{20}$~cm$^{-2}$  for $E\tief{B-V}=0\fm05$).
 We interpret
the large hydrogen column densities as evidence for the symbiotic
nebula being radiation bounded.

The strength of the  \1640 line is a measure for the number of   stellar photons
emitted below $\lambda=228$~\AA, but weighted strongly for photons just
shortwards of the threshold wavelength. Thus, basically, we use
a sophisticated Zanstra method. As 
it is the case for the original Zanstra method, our
analysis depends 
only weekly on the adopted parameters of the nebula. 
A comparison of the calculated \1640
line flux with the observed one selects those models with the correct
number of He$^+$ ionizing photons. 

For a black body energy distribution 
the \1640 analysis yields a unique  temperature. 
If we use 
emergent radiation from hydrostatic atmospheres, the \1640 analysis selects
different effective temperatures depending on the chemical composition
of the atmosphere and -- to a much lower degree --
on the gravity. 
However, although there is a range of temperatures of plane-parallel models 
that fits the observed \1640 flux, the \1640 analysis still determines
the stellar temperature reasonably well.

\begin{figure}[t]
\epsfxsize=0.5\textwidth
\epsffile{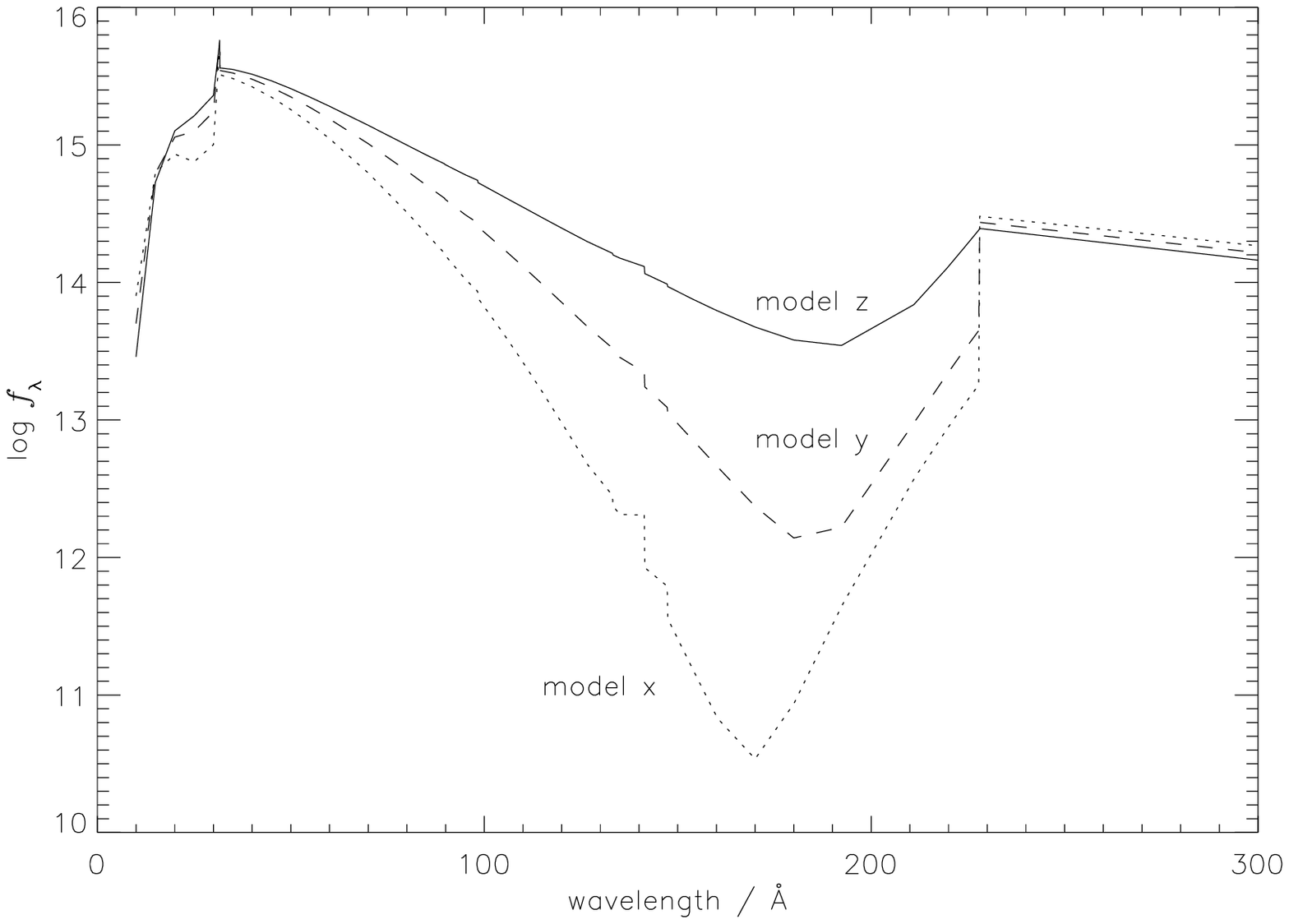}
\caption[]{Relative energy distributions of three models with
approximately the same effective temperature of \Teff$ = 300$ kK
but with different mass loss rates 
(model \underline{x}: $\Mdot_{\underline{x}}=10^{-5.1}$\,\Msolar/yr,
      \underline{y}: $\Mdot_{\underline{y}}=10^{-5.2}$\,\Msolar/yr, and 
      \underline{z}: $\Mdot_{\underline{z}}=10^{-5.3}$\,\Msolar/yr;
see Table\,\ref{wrmod} for the other parameters). Note that a final
velocity of $v_\infty=1000$\,km/sec is assumed (only $\Mdot/v_\infty$
enters as a parameter into the model calculation)
}
\label{mloss}
\end{figure}

The results using atmospheres with a stellar wind 
differ fundamentally from those based on plane-parallel models. 
If we feed the model flux 
resulting from our calculations with extended model atmospheres with mass loss 
to the \1640 analysis,  we find that
the number of He$^+$ ionizing
photons is a sensitive function mainly of the mass loss, and not
of the stellar temperature (see Sect.\,\ref{swind}).
The flux below 228 \AA\ is suppressed by absorption in the wind depending
on the mass loss rate. If the mass loss rate is sufficiently large
 helium recombines to He$^+$ and in this case, the wind is
optically thick to photons more energetic than the He$^+$ edge
 (Schmutz et al. 1992).
In Fig.~\ref{mloss} we show the EUV and soft X-ray energy distribution 
predicted by three  models with about the same 
effective temperature of  \Teff$=300,000$ K but
 different mass loss rates 
(models \underline{x} -- \underline{z}). 
Thus, for wind models the \1640 analysis does not provide information on
the effective temperature except that there is a lower limit in order to
produce He$^+$ ionizing radiation. However,
 no upper limit to the effective temperature of the star can be specified:
We can always increase the
mass loss rate to reduce the number of He$^+$ ionizing photons to the 
value implied by the observed \1640 flux, no matter what stellar temperature
we choose.

The third step of our analysis is to compare the X-ray flux of the
model atmosphere with the flux observed by ROSAT. For this
comparison the theoretical spectrum has to be attenuated to account
for the absorbing interstellar and circumstellar material. We calculate the
absorption  according to
Morrison \&\ McCammon (1983) and Cruddace et al. (1974).
The attenuated spectrum is then
converted to a pulse height distribution (PHD) by convolution  with
the PSPC detector response matrix  and the effective areas provided 
by the EXSAS package. 
The comparison of the
theoretical PHD 
with the low energy part of the ROSAT observation, 0.1 to 0.2 keV, 
allows to determine the hydrogen column density. The comparison of the
high energy part tests the emergent flux of our models at 
wavelengths where the flux is sensitive to the stellar temperature and to
the presence of metals (e.g. the C$^{+4}$ ionization edge at 0.39\,keV).

\def\xx{[$10^{-5}$]}

\begin{table*}[t]
\caption{Summary of model atmospheres parameters. In
each series
we computed atmospheres of several temperatures until a fit to the
UV continuum and the \ion{He}{ii} emission could be achieved. 
The chemical abundances are given as number fractions.
}
\label{wrmod}
\begin{flushleft}
\begin{tabular}{llrllrrrrrl}
\hline
Series & Model type &$\log\Mdot/v_{\infty}$ 	&$\log g$	&$\beta$
	&H &He &C &N &O	&Abundances \\
&	&[\Msolar/(yr km\,sec$^{-1})$]&	&	&  &  &\xx &\xx &\xx &similar to \\
\hline
\underline{A} & hydrostatic	&-& 5 -- 7	&-&$0.90	$
		&$0.10	$&$48.00$&$10.00	$&$84	$&Sun\\
\underline{B} & hydrostatic	&-	&6	&-&$0.90	$
		&$0.10	$&$4.80	$&$1.00	$&$8.4	$&Magellanic Clouds\\
\underline{C} & hydrostatic	&-	&7.5	&-&$1.00	$
		&$0.00	$&$0.00	$&$0.00	$&$0.00$&DA\\
\underline{D} & hydrostatic	&-	&6 -- 7.5&-&$0.01	$
		&$0.99	$&$100.00		$&$0.00	$&$0.00	$&DO\\
\underline{E} & hydrostatic	&-	&6 -- 7	&-&$0.01	$
		&$0.99	$&$0.10	$&$0.00	$&$0.00	$&DO\\
\underline{F} & hydrostatic	&-	&5.5 -- 6&-&$0.67	$
		&$0.33	$&$10.00$&$0.00	$&$0.00	$&DO\\
\underline{G} & hydrostatic	&-	&6	&-&$0.50	$
		&$0.50	$&$100.00$&$0.00	$&$0.00	$&DO\\
\underline{H} & hydrostatic	&-	&6 -- 7	&-&$0.50	$
		&$0.50	$&$10.00	$&$0.00	$&$0.00	$&DO\\
\hline
\underline{p} & WR-type  	&$-8.53$ &-&1
       		&0.50		&0.50 	&$40.00$& $0.76$ &$7.50$ &-\\
\underline{q}& WR-type 	&$-8.50$      &-&1
       		&0.50  &0.50 	&$40.00$ 	&$0.76$ &$7.50$ &-\\
\underline{r}& WR-type 	&$-8.12$  &-&1 
       		&0.50	&0.50 	&$100.00$ &$0.76$&$100.00$ &-\\
\underline{s}& WR-type	&$-7.97$              &-&1
       		&0.50	&0.50 	&$200.00$ &$0.76$&$100.00$ &-\\
\underline{t}& WR-type	&$-8.30$              &-&1
       		&0.50	&0.50 	&$20.00$ &$0.76$&$7.50$ &-\\
\underline{u}& WR-type	&$-8.30$              &-&1
       		&0.50	&0.50 	&$40.00$ &$0.76$&$7.50$ &-\\
\underline{x}& WR-type   &$-8.10$              &-&1
       		&0.50	&0.50 	&$40.00$ &$0.76$&$7.50$ &-\\
\underline{y}& WR-type   &$-8.20$              &-&1
       		&0.50	&0.50 	&$40.00$ &$0.76$&$7.50$ &-\\
\underline{z}& WR-type 	&$-8.30$  &-&1
       		&0.50    &0.50     &$40.00$ &$0.76$&$7.50$ &-\\
\hline
\end{tabular}
\end{flushleft}
%
\caption{Solutions to the UV continuum and \1640
 emission. The $5^{\rm th}$
column tells us whether a fit to the ROSAT
observation could be achieved. The comment `too soft' means that
too much counts are predicted at low energies compared to the observation.
Column 6 lists the implied interstellar hydrogen  column density. }
\label{solu}
\begin{flushleft}
\begin{tabular}{llrrlrr}
\hline
Model & Model type &\Teff	&$L_\star$	
	&Comment on implied ROSAT solution	
		&$\Nh$	&Fig.	\\
&&[kK] 	&$[\Lsolar]$&&[$10^{20}$cm$^{-2}$]&\\
\hline
Black Body & Black Body
 &100 &8\,100  
		&count rate is by far too low		& $<0.1$&-\\
\hline
\underline{A}& hydrostatic&100 &11\,000 	
		&count rate is too low				&$0.0$&-\\
\underline{B}& hydrostatic&120 &17\,000  	
		&synthetic PHD is too soft			&$0.1$	&-\\
\underline{C}& hydrostatic&65&5\,800
		&synthetic PHD too low at 0.3--0.4\,keV		&$0.4$	&-\\
\underline{D}& hydrostatic&130 &23\,000  	
		&synthetic PHD too low at $E \ga 0.25$\,keV	&$4.0$	
&\ref{fdo}\\
\underline{E}& hydrostatic&120 &17\,000	
		&synthetic PHD too low at $E \ga 0.20$\,keV	&$1.7$	
&\ref{fdo}\\
\underline{F}& hydrostatic&120 &17\,000  	
		&synthetic PHD too low at $E \ga 0.20$\,keV	&$2.0$	& \\
\underline{G}& hydrostatic&120 &16\,000  	
		&synthetic PHD too low at $E \ga 0.20$\,keV	&$2.1$	& \\
\underline{H}& hydrostatic&120 &17\,000  	
		&synthetic PHD too low at $E \ga 0.20$\,keV	&$2.6$	& \\
\hline
\underline{p}&WR-type&220 &$9\,200$ 
		&PHD slightly too low at 0.20--0.45\,keV &$7.7$	&
\ref{wr748690} \\
\underline{q}&WR-type&265 &$10\,200$
		&PHD slightly too low at 0.30--0.45\,keV		&$8.7$	&4\\
\underline{r}&WR-type&309 &$11\,200$ 
		&relative good fit		&$8.8$	&
\ref{wr748690}\\
\underline{s}&WR-type&354 &$12\,700$ 
		&relative good fit		&$9.3$	
&\ref{wr748690}\\
\underline{t}&WR-type&300 &$16\,800$ 
		&PHD too high  at $E>0.45$\,keV		&$8.8$	
&\ref{fwr7071}\\
\underline{u}&WR-type&300 &$16\,800$
		&PHD slightly too low at 0.30--0.42\,keV	&$8.8$	
&\ref{fwr7071}\\
\underline{x}&WR-type&315 &$11\,500$
		&PHD too high  at $E>0.37$\,keV		&$11.0$	&
\ref{mloss},\ \ref{fwr77_78_79}\\
\underline{y}&WR-type&325 &$16\,000$
		&PHD too high  at $E>0.35$\,keV		&$11.0$	&
\ref{mloss},\ \ref{fwr77_78_79}\\
\underline{z}&WR-type&335 &$22\,900$
		& PHD too high  at $E>0.33$   &$11.0$   & 
\ref{mloss},\ \ref{fwr77_78_79}\\
\hline
\end{tabular}
\end{flushleft}
\end{table*}

\section{Results}
\label{res}\label{resu}

In Table~\ref{wrmod} we list those   parameter ranges for which we
have calculated model atmospheres and for which the prediced flux distribution
is able to 
reproduce the observed continuum flux at 1300 \AA\ as well as the \1640 line
flux. For these models the predicted X-ray luminosity was compared
to the ROSAT observations. In Table~\ref{solu} we summarize
the results of these fits, and list the resulting \Teff\ and $L_*$, and
for the models with stellar wind $\Mdot/v_\infty$. Also shown is 
the hydrogen column density required to fit the low energy X-ray part.
In principle we could quantify the quality of the fits by listing the
reduced $\chi^2$ of the solution. However, one can  clearly see  that the
deviations between the predictions and the observations are not solely due to
statistical errors: Even in the case of the  ``relative good fits'' systematic
deviations remain which may have its origin in small incompleteness
of the modelling. 
Finally,  Table~\ref{wrmod}  comments on the quality  of the resulting fit to
the ``high''-energy tail X-ray flux.

\subsection{Black body emission}

If we assume that the radiation of the hot component has a
black body spectral distribution
the \1640 analysis yields a unique solution, namely
$T_{\rm BB} = 100,000$\,K and $R = 0.3$\,\Rsolar.
However, the emergent X-ray flux of this solution is 
much
too low to reproduce
the PSPC count rate, even for  a vanishing interstellar hydrogen column
density. Moreover, no [\ion{Fe}x] emission is expected
because the radiation field is not hard enough to produce Fe$^{+9}$.

Good fits to the PHD can be achieved if either the stellar radius
is increased to about $1.5 \cdot 10^7$\,\Rsolar\ or the temperature is
increased to $T\tief{BB}\ga220,000$\,K (with $R\approx 2~\Rsolar$),
in agreement with the result seen in Fig.\,1 of Kahabka et al. (1994).
On obvious reasons, these parameters clearly lead to an UV flux 
several orders of magnitude too high 
compared to the observation. Also, the luminosity would exceed
the Eddington limit by many orders of magnitude.

\begin{figure}[htbp]
\epsfxsize=0.5\textwidth
\epsffile{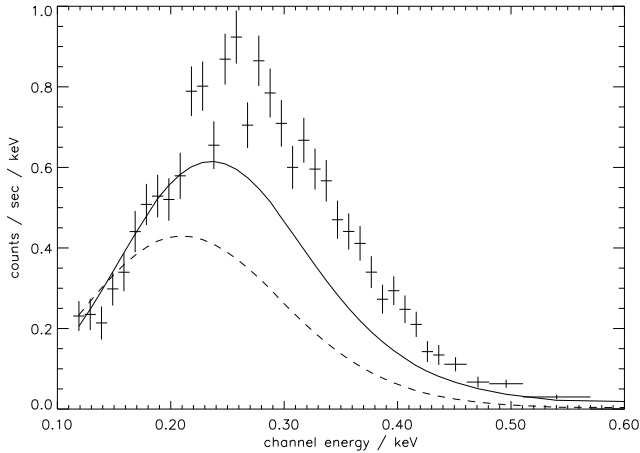}
\caption[]{The solid line is the best solution to the UV and X-ray observations
with DO model atmospheres
(model \underline{D} in Tab.\,\ref{wrmod}):
${\rm He/H} = 100$, ${\rm C/H} = 0.1$,
$\Teff = 130,000$\,K, and $\Nh = 4.0\cdot10^{20}\,{\rm cm}^{-2}$.
Also shown is one example for another chemical composition (dashed line,
model \underline{E}):
${\rm He/H} = 100$, ${\rm C/H} = 10^{-4}$,
$\Teff = 120,000$\,K, and $\Nh = 1.7\cdot10^{20}\,{\rm cm}^{-2}$. The other
compositions result in nearly the same synthetic PHDs}
\label{fdo}
\end{figure}

\subsection{Hydrostatic models}

Models with solar abundances (series \underline{A}, see Tab.~3)
lead to results similar to the  black body
case: the UV measurements would result in $\Teff \approx 100,000$\,K whereas
$\Teff \ga 200,000$\,K is needed to reproduce the X-ray observations.
If we reduce the CNO
abundances by a factor $10$ (series \underline{B}), the UV measurements are
in agreement with
the predictions at $\Teff\approx$\,120,000\,K.
 However, this is still too cold to account for the X-ray flux.

Pure hydrogen (DA, series \underline{C}) atmospheres produce by far the
``hardest'' possible spectrum, i.e. the highest X-ray flux for a given
temperature. Again, these models lead to the situation that 
a relatively good fit to the PSPC PHD can be obtained at
$\Teff \approx 100,000$\,K, 
but the 
\1640 line flux is only reproduced at $\Teff \la 65,000$\,K.

Next, we assumed  that helium is the main constituent of the hot component's
atmosphere.  We examined DO/DOA model atmospheres with different
helium-to-hydrogen ratios.
The best solution to both, the UV and X-ray measurements is obtained
with ${\rm He/H} = 100$, ${\rm C/H} = 0.1$, and $\Teff = 130,000$\,K
(Fig.\,\ref{fdo}, model \underline{D}). If the hydrogen
column density is chosen so that the lowest channels of the pulse height
distribution are reproduced then the predicted flux is somewhat too small
at higher energies. This deviation is larger for all other combinations
of helium and carbon abundances (see Fig.\,\ref{fdo}).
In the atmosphere of a hot DO white dwarf with $\Teff > 100,000$\,K
nitrogen, oxygen, and iron group elements may
also be present. However, if we assume metal abundances as high as in
PG\,1034+001 ($\Teff \approx 100,000$\,K, Werner et al. 1995a)
the X-ray flux is completely absorbed.

Therefore an ionized spectrum is required
\begin{itemize}
\item
with a strong \ion{He}{ii}$\;\lambda228$
absorption edge in order to prevent moderately ionized lines like \1640 from
becoming too strong.
\item
and a considerable portion of the energy  released beyond the
``gap'' due to \ion{He}{ii} and possibly metals,  to absorb high-energy photons.
\end{itemize}

\subsection{Models with a stellar wind}
\label{swind}

If we assume that the hot object in SMC3 has a stellar wind  we
have to determine three model parameters, one more -- the mass loss rate -- 
in addition to effective temperature and stellar radius. As pointed
out in Sect.~\ref{sfit}, the stellar wind allows to suppress the He$^+$
ionizing photons even for very high stellar temperatures. With three model 
parameters --- \Teff, $R_*$, and $\Mdot/v_{\infty}$ --- and three
observational constraints --- the continuum flux at 1300 \AA, the flux
of the \1640 line, and the X-ray emission --- it should be possible,
in principle, to determine a solution.

However, in practice, we find that no unique solution exists. 
The problem is that
if we fit both UV constraints, the 1300~\AA\
continuum and the \1640 line flux by adjusting two of the three parameters,
say $R_*$, and $\Mdot/v_{\infty}$ as a function of \Teff,  we find that the 
predicted X-ray flux depends on the third parameter, \Teff, only in a
limited range of temperatures:  In
particular, we find that for lower temperatures ($\approx 100-200$\,kK)
the X-ray flux is quite sensitive to \Teff; at higher temperatures, however,
a saturation effect occurs and we obtain  an upper limit to the
predicted X-ray luminosity, almost regardless of the specified effective
temperature.

It turns out that this upper limit reproduces almost, but not perfectly,
the observed X-ray spectrum. The reason for this result is a typical property
of a stellar wind.
The wind that suppresses the photons at the
the He$^+$ ionization edge is still relatively opaque in the observed X-ray
region below 100 \AA. This is simply because the bound-free continuum
cross section is proportional to $\lambda^{3}$. Therefore, as long
as we have a stellar temperature where the maximum of the radiation is 
in the observed X-ray or beyond, 
we find that all models have similar X-ray energy distribution determined by
the helium absorption cross section. This is similar to adjusting the
interstellar column density \Nh. All these arguments are only valid for the
energy spectrum around 100 \AA (where the bulk of the observed X-ray flux
originates). The high energy tail, where the influence on the helium 
bound-free absorption becomes smaller, would in principle be  very sensitive
to the temperature. However, as discussed below, the presence of  carbon
absorbs all flux below 31.6\,\AA.

\begin{figure}[htbp]
\epsfxsize=0.5\textwidth
\epsffile{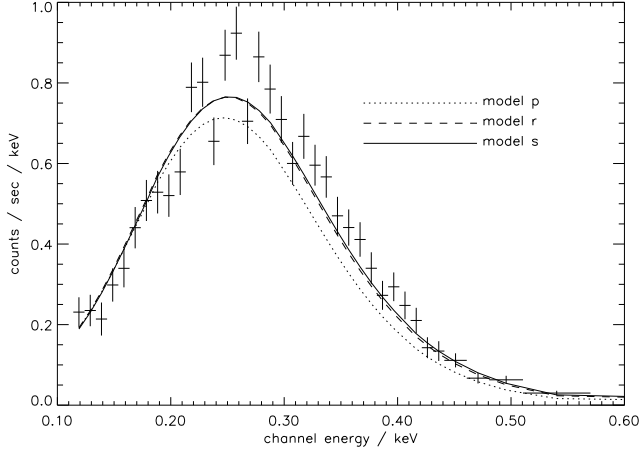}
\caption[]{
Comparison of the observed PHD of SMC3 with 
three  Wolf-Rayet type models
 with
$\Teff = 220$~kK (model \underline{p}, see Table\,\ref{wrmod} for the other
parameters), 309~kK (\underline{r}), and 354\,kK (\underline{s}).
The predicted flux from model \underline{p} is not hard enough to reproduce
the observation, while the two hotter models (\underline{r}) and 
(\underline{s}) lead to relatively good  fit to the data
}
\label{wr748690}
\end{figure}

Our difficulty to determine a unique set of the three parameters
\Teff, $R_*$, and $dM/dt$ does not prevent
us to determine the stellar luminosity. 
We have stated above that the form of the energy
distribution is well determined and therefore, since we know the
distance to our object, we know the luminosity 
with only a weak dependence on the other parameters. 
The statement that the effective temperature is not 
well known, is physically speaking a limit in determining the 
hydrostatic radius of the hot object, because this radius is hidden
by the stellar wind. From models \underline{r} and 
\underline{s} (see Table~\ref{solu}, Fig.\,\ref{wr748690}) that yield
the best fit to  the observations, we 
find $L_{\mbox{SMC3}}=10^{4.05\pm 0.05}$\,L$_\odot$.

\begin{figure}[htbp]
\epsfxsize=0.5\textwidth
\epsffile{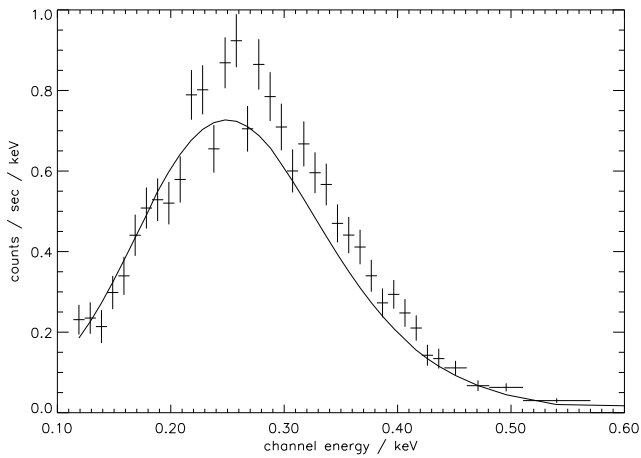}
\caption[]{
The ``coolest''   Wolf-Rayet type model with \Teff=260\,K 
(\underline{q}) that  is able to reproduce the observed ROSAT PHD
reasonably well
}
\label{fwrq}
\end{figure}

\begin{figure}[htbp]
\epsfxsize=0.5\textwidth
\epsffile{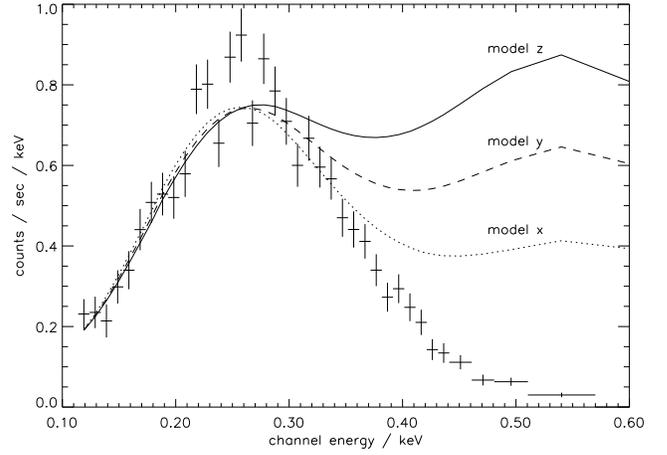}
%
\caption[]{Predicted ROSAT PHD for the models with different mass-loss
rates  shown in Fig.\,\ref{mloss}: $\Mdot_{\underline{x}}=10^{-5.1}$\,\Msolar/yr,
$\Mdot_{\underline{y}}=10^{-5.2}$\,\Msolar/yr, and
$\Mdot_{\underline{z}}=10^{-5.3}$\,\Msolar/yr;
$v_\infty=1000$\,km/sec is assumed
}
\label{fwr77_78_79}
\end{figure}

\begin{figure}[htbp]
\epsfxsize=0.5\textwidth
\epsffile{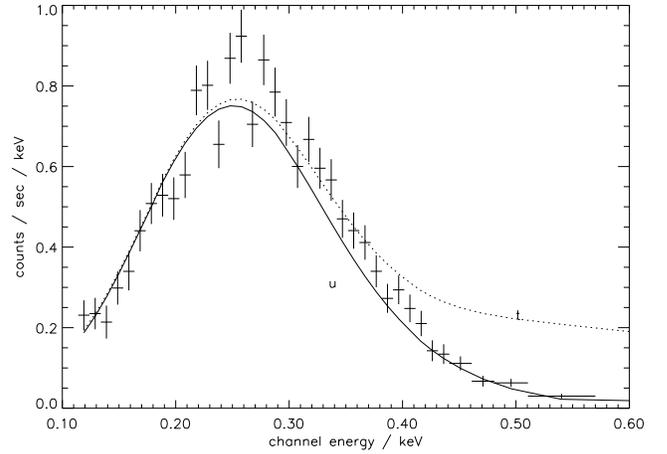}
%
\caption[]{These wind models  with $\Teff\approx 270$\,kK 
(model \underline{t} and \underline{u}) show,  that too much hard
 X-ray photons
are predicted if the carbon abundance is too low 
(C/He=$2\cdot 10^{-4}$ in the case of \underline{t}
compared to  and $4\cdot 10^{-4}$ for \underline{u})
}
\label{fwr7071}
\end{figure}

Although we cannot decide which temperature is to be preferred, 
$\Teff= 309$\,kK, 354\,kK, or even hotter, we can
set a lower limit to the effective temperature,
i.e. an upper limit to the stellar radius.
In Fig.~\ref{wr748690} 
we show fits to the observed X-ray data based on models with effective
temperatures of $\Teff = 220$~kK (model \underline{ p}),
309~kK (\underline{r}), and 354\,kK (\underline{s}). 
It is evident that
the radiation of the model \underline{p}  is not quite hard
enough
to reproduce the observed pulse height distribution.
On the other hand, both hotter models
come close to fitting the observed X-rays.
Fig.\,\ref{fwrq} shows that  a
model with $\Teff = 265$\,kK (model \underline{q}) 
is the coolest to
fit  the energy distribution of the hot object in SMC3 reasonably well.

In order to derive $\Mdot$ a value for $v_\infty$ is required.
Unfortunately, $v_\infty$ is not easily measurable. We do not
dispose of the high S/N, high resolution UV spectra which allow
to detect broad line components and determine their width.
Even for galactic objects the velocity of the winds from
hot components of symbiotic systems could be measured only
for a few objects. Nussbaumer et al.\ (1995) and
Nussbaumer \& Vogel (1995) derive $v_\infty\approx1000$~km/s
which we will adopt in the following for SMC3. With this value,
model \underline{q} implies a rather
high mass loss rate of $\Mdot>10^{-5.3}$~M$_\odot$\,yr$^{-1}$.

Helium and carbon are the only elements that have a
significant influence on the
resulting spectrum. Helium is needed in order to produce the
absorption at the He$^+$ edge. The mass loss rate quoted above is
basically the mass loss of helium. We have assumed a hydrogen abundance
of 20 \%\ by mass, but the hydrogen abundance could be changed 
arbitrarily without observable consequences. The carbon 
abundance is found to be very important. With the hot temperatures of our
best fitting models we predict significant flux shortwards of the
C$^{+4}$ ionization edge at 31.6~\AA\ (0.39\,keV).
Such energy distributions would produce much larger count rates 
than observed at energies larger than 0.4\,keV.
Therefore a relatively high carbon abundance is required
in order to fit the observed energy distribution.
 Basically, we are invoking the same wind absorption
effect at the C$^{+4}$ edge as we invoked above to reduce the numbers of 
He$^+$ ionizing photons. 
Fig.\,\ref{fwr7071} demonstrates the importance of the carbon edge by comparing
the predicted ROSAT PHD for C/He=$2\cdot 10^{-4}$ (too much ``hard'' X-ray
flux) and $4\cdot 10^{-4}$ (slightly too low flux at $E>0.3$\,keV)
for about the same effective temperature ($\approx 270$\,kK).
 A slightly larger relative carbon abundance
is required to suppress the high energy photons for the hotter models. 
Therefore this carbon abundance is a minimum value and we cannot exclude 
higher values.

\section{Conclusions\label{conclu}}

SMC3 is extraordinary, even among symbiotic stars.
It is remarkably X-ray bright and shows unusually 
high ionization in the nebula. We succeeded to model these
properties and the UV flux with a very hot WR-type model atmosphere.
Our result clearly confirms that at least some of the 
supersoft X-ray sources contain very hot photospheres of degenerate
stars. It shows in particular, that such hot stars can be encountered 
in symbiotic binary systems. Reasonable fits are achieved  at
temperatures above $260\,000$~K and mass-loss rates of 
$\Mdot\gappr1\cdot10^{-6}$~\Msolar/yr, (assuming $v_\infty=1000$~km/sec).
Due to the properties of the wind
models (the hot star itself is hidden by the stellar wind)
no  upper limit for the effective temperature could be specified.
However, this uncertainty does not effect the total luminosity:
$L_{\mbox{SMC3}}=10^{4.05\pm 0.05}$\,L$_\odot$.

Even the lower limit for the effective  temperature is extraordinarily high
(see the temperature distribution for hot components of symbiotic
systems in Fig.~7 of Paper III). To our knowledge SMC3 has the hottest stellar
 atmosphere ever derived in 
a sophisticated NLTE analysis. Until now, the hottest known \hbox{(pre-)} white 
dwarfs are the DA central star of the planetary nebula WDHS1 with
\Teff=160\dots 200\,kK (Liebert et al. 1994; Napiwotzki 1995)
and the PG1159 stars H1504+65 (Werner 1991) and RXJ2117+3412
(Werner et al. 1995b) with \Teff=170\,kK. Note, however, that the temperature
of SMC3 refers to the hydrostatic radius $R_*$ and not to $\tau_{R}=2/3$.

On the basis of LTE model atmospheres Heise et al. (1994) have shown that
the ROSAT PSPC observations of the  super-soft X-ray source 1E0056.8-7154,
associated with the planetary nebula N67 (Wang 1991; Cowley et al. 1995)
can be explained assuming a hot pre-white dwarf with an
effective temperature as high as 450\,kK.
However, at such high temperatures NLTE effects are expected to be strong
so that this result has to be checked using NLTE model atmospheres.

Strong mass loss is indispensible to explain the spectrum
of SMC3. Direct or indirect signatures of strong mass loss 
are also encountered in other symbiotic novae
(see Sion et al.\ 1993, JMW, Nussbaumer et al.\ 1995, 
Nussbaumer \& Vogel 1995, and the discussion in M\"urset et al.\ 1995a).
If SMC3 has been shedding its wind constantly since the
beginning of the outburst, it must have lost at least
a mass of $\Delta M\sim10^{-4}$~M$_\odot$. This is much
larger than the mass burnt during the same time, which means
that the duration of the outburst of SMC3 will be limited 
rather by the mass loss than by the nuclear processes.
$\Delta M$ is also a lower limit to the mass of the
layer accreted by the white dwarf prior to outburst.

Kenyon et al.\ (1993),
M\"urset \& Nussbaumer (1994), and JMW found for the known 
galactic symbiotic novae luminosities clearly below the
Eddington limit. This result is now confirmed with a result
that, due to the well known distance, is probably
more reliable than the galactic ones. Here we encounter a basic 
difference between symbiotic and classical novae
which seem to exceed the Eddington limit
(Starrfield et al.\ 1993; Shore, personal communication). 
If a core-mass -- luminosity  relation holds for the
symbiotic novae we can derive a mass $M\approx0.8$~\Msolar for 
the hot star (with the relation from Joss et al.\ 1987).

As a final comment we would like to stress that a photospheric 
X-ray source like SMC3 can only be reasonably investigated
with i) multi-frequency data and ii) sophisticated 
\underline{non}-LTE atmosphere models. Kahabka et al.\ (1994)
impressively demonstrated the failure of black body fits
to the ROSAT PHD by deriving a luminosity of $L\sim10^{13}$~L$_\odot$
for SMC~3. The reasons for the failure are obvious:
Fitting the ROSAT data alone is a dangerous extrapolation, 
because they represent only an extreme tail of the stellar
spectrum. Even worse, this extreme part is exactly the 
spectral region where hot atmospheres differ very strongly
from Planck curves.

\acknowledgements
We thank Steven N.~Shore for helpful comments.
Work on ROSAT data in Kiel is supported by DARA grant 50 OR 94091.
KW acknowledges support by the DFG under grant \hbox{We1312/6-1}.

\end{document}